\documentclass[aps,prb,amsmath,amssymb,superscriptaddress,reprint,
]{revtex4-2}
\usepackage{amsmath}

\usepackage{graphicx}
\usepackage{dcolumn}
\usepackage{bm}
\usepackage{hyperref}
\hypersetup{colorlinks=true,allcolors=blue}

\usepackage{siunitx}
\usepackage{physics}
\usepackage{color}
\usepackage{mathptmx}

\begin{document}

\title{Spin-polarized transport properties in magnetic moiré superlattices} 
 
\author{Zhao Gong}
\affiliation{Hebei Provincial Key Laboratory of Photoelectric Control on Surface and Interface, School of Science, Hebei University of Science and Technology, Shijiazhuang, 050018, China}

\author{Qing-Qing Zhang}
\affiliation{Department of Physics and Information Engineering, Cangzhou Normal University, Cangzhou 061001, China}

\author{Hui-Ying Mu}
\email[Correspondence to: ]{ xiaomu1982@163.com}
\affiliation{Hebei Provincial Key Laboratory of Photoelectric Control on Surface and Interface, School of Science, Hebei University of Science and Technology, Shijiazhuang, 050018, China}

\author{Xing-Tao An}
\email[Correspondence to: ]{ anxt2005@163.com}
\affiliation{Hebei Provincial Key Laboratory of Photoelectric Control on Surface and Interface, School of Science, Hebei University of Science and Technology, Shijiazhuang, 050018, China}
\affiliation{College of Physics, Hebei Normal University, Shijiazhuang 050024, China}
\affiliation{Key Laboratory for Microstructural Material Physics of Hebei Province, School of Science, Yanshan University, Qinhuangdao 066004, China}

\author{Jian-Jun Liu}
\affiliation{College of Physics, Hebei Normal University, Shijiazhuang 050024, China}
\affiliation{Department of Physics, Shijiazhuang University, Shijiazhuang 050035, China}

\date{\today}

\begin{abstract}

Since the discovery of the fascinating properties in magic-angle graphene, the exploration of moiré systems in other two-dimensional materials has garnered significant attention and given rise to a field known as “moiré physics”. Within this realm, magnetic van der Waals heterostructure and the magnetic proximity effect in moiré superlattices have also become subjects of great interest. However, the spin-polarized transport property in this moiré structures is still a problem to be explored. Here, we investigate the spin-polarized transport properties in a moiré superlattices formed by a two-dimensional ferromagnet CrI$_{3}$ stacked on a monolayer BAs, where the spin degeneracy is lifted because of the magnetic proximity effect associated with the moiré superlattices. We find that the conductance exhibits spin-resolved miniband transport properties at a small twist angle because of the periodic moiré superlattices. When the incident energy is in the spin-resolved minigaps, the available states are spin polarized, thus providing a spin-polarized current from the superlattice. Moreover, only a finite number of moiré period is required to obtain a net spin polarization of 100\%. In addition, the interlayer distance of the heterojunction is also moiré modifiable, so a perpendicular electric field can be applied to modulate the direction of the spin polarization. Our finding points to an opportunity to realize spin functionalities in magnetic moiré superlattices.

\end{abstract}

\maketitle

\sloppy

\section{\label{sec:Inro}Introduction}

The discovery of van der Waals moiré superlattices in twisted bilayer graphene has revealed a plethora of unique phenomena, including unconventional superconductivity \cite{cao2018unconventional,yankowitz2019tuning}, ferromagnetism and quantum anomalous Hall effect \cite{sharpe2019emergent,serlin2020intrinsic}, correlated insulating states \cite{cao2018correlated}, and ferroelectricity \cite{vizner2021interfacial}, among others. The concept of magic-angle graphene has also been extended to other two-dimensional (2D) materials with similar twisted angle structures. For example, in WS$_{2}$/WSe$_{2}$ heterostructures, a Mott insulating state with antiferromagnetic ordering is discovered when the valence band topological superlattice is integer-filled \cite{regan2020mott,tang2020simulation}. Additionally, there exist certain special correlated insulating states under other fractional fillings \cite{xu2020correlated}. Evidence of moiré excitons has also been observed in heterostructures formed by other transition metal dichalcogenides (TMDs) \cite{tran2019evidence,seyler2019signatures,alexeev2019resonantly,jin2019observation}. 

The experimental success of magic-angle graphene is also attributed to the early theoretical predictions \cite{bistritzer2011moire}. In a moiré superlattice, the local atomic registry and interlayer interactions undergo a slowly varying periodic modulation which can be described by a moiré potential \cite{yu2017moire,tong2019magnetic}. The concept of moiré potential has found successful applications in the theoretical predictions of moiré excitons \cite{yu2017moire}, quantum spin Hall effect \cite{wu2019topological,yu2020giant,zhai2022ultrafast,tong2017topological}, correlated insulating states \cite{wu2018hubbard} and chiral phonons \cite{yu2022phonons}.

In recent years, magnetic 2D materials and their moiré physics have also attracted considerable attention. In monolayer, bilayer, trilayer, and under twist angles, CrI$_{3}$ exhibits a lot of intriguing magnetic orders and magnetic domains \cite{huang2017layer,wang2020stacking,song2021direct,sivadas2018stacking,akram2021moire}. In the heterostructures formed by stacking magnetic materials and non-magnetic semiconducting 2D materials, a magnetic proximity effect exists, leading to the spin splitting and valley splitting in the non-magnetic 2D materials \cite{seyler2018valley,voroshnin2022direct,norden2019giant,tang2020magnetic,zhong2020layer}. For example, it has recently been discovered that a magnetic proximity effect can be detected in the ferromagnetic material/graphene system and can be designed for spin transport devices \cite{tang2020magnetic,zhao2023room}. This notion appears to be particularly captivating within the moiré system, as exemplified by the presence of a moiré modifiable magnetic proximity effect in the CrI$_{3}$/BAs moiré superlattice system \cite{tong2019magnetic}. Due to the influence of periodic moiré potential, BAs exhibits the localized miniband effect in electron behavior.

\begin{figure*}
	\centering
	\includegraphics[width=17cm]{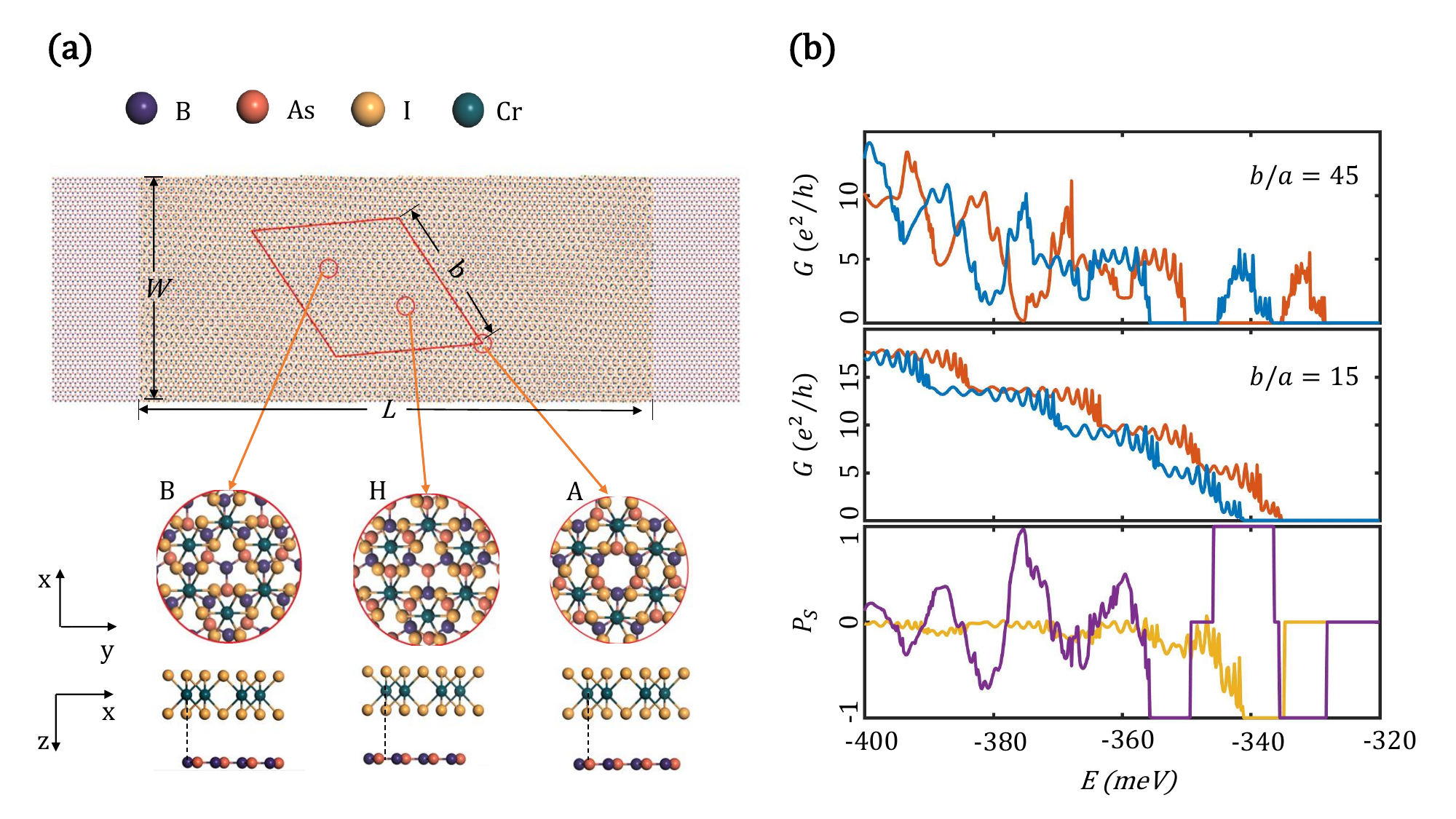}
	\caption{\label{fig:a1}The miniband transport and spin polarization in a moiré superlattice. (a) The schematic shows a superlattice nanoribbon, where the central scattering region of a long strip of monolayer BAs nanobelt is covered by CrI$_{3}$ to form moiré patterns. The atomic stacking between heterojunctions smoothly varies to create three high-symmetry points. (b) The spin-dependent conductance and spin polarization as the functions of the incident energy for moiré period of $b/a=45$ and $b/a=15$. The red and blue lines represent the conductance of spin-up and spin-down electrons, respectively. The purple and yellow lines represent the spin polarization at $b/a=45$ and $b/a=15$, respectively.
	}
\end{figure*}

In this paper, we investigate the transport properties of the nanoribbon devices based on a moiré superlattice formed by stacking a monolayer of ferromagnetic material (CrI$_{3}$) on a monolayer of semiconductor (BAs). Due to the lattice mismatch and twist angle, such magnetic van der Waals heterostructures give rise to moiré patterns, and the interlayer interactions strongly depend on the local interlayer atomic registry. As a result, the local interlayer interaction, magnetic proximity effect, and interlayer distance can all be modulated by the moiré effect. It is possible to modulate the spin splitting by applying a perpendicular electric field \cite{tong2019magnetic}. The moiré-modulated interlayer interaction and magnetic proximity effect can be described by a spin-dependent moiré potential, while the interlayer distance-dependent modulation by the perpendicular electric field can be described by a spin-independent moiré potential.

Periodic moiré superlattices exhibit the spin-dependent minibands transport properties, and this effect is closely related to the moiré period, which can be modulated by the twist angle. The calculated results indicates that larger moiré periods can lead to flatter and narrower minibands, which eventually result in larger spin-dependent minigaps. The spin polarization becomes most prominent when the minigap modulation approaches the magnitude of the spin splitting. Furthermore, due to the presence of spin splitting, the response of the spin-up and spin-down conductance to the electric field is different, enabling the direction of spin polarization achieved effective modulation through applied electric fields. Our calculations demonstrate that spin polarization is symmetric within a certain range of electric field modulation. The finding provides a routing for the manipulation of spin degrees of freedom and sheds light on the theoretical study and design of spintronic devices.

\section{\label{sec:theor}MODEL AND METHODS}

We consider a zigzag monolayer BAs nanoribbon, where a segment with length ${L}$ and width ${W}$ is covered by a monolayer of CrI$_{3}$, acting as the central scattering region, as shown in the upper panel of Fig.~\ref{fig:a1}(a). Due to the presence of twisting and lattice mismatch in such bilayer heterostructures, a periodic moiré superlattice is formed. The moiré period of a superlattice can be expressed by the formula $b \approx {a}/{\sqrt{\delta^2 + \theta^2}}$, where $\mathrm{a}$ is the lattice constant of BAs, $\delta$ represents the lattice mismatch, and $\theta$ represents the twist angle between the two layers.

The valence band edge of monolayer BAs is located within the band gap of CrI$_{3}$, approximately $450 meV$ away from the conduction band edge of CrI$_{3}$ and around $700 meV$ away from the valence band edge of CrI$_{3}$. We focus only on the energy range within $80 meV$ of the valence band maximum of BAs. As a result, electrons within these energy ranges do not undergo interlayer transitions. Additionally, the strong spin-orbit coupling in  twisted bilayer CrI$_{3}$ plays a crucial role in its magnetism\cite{sivadas2018stacking,akram2021moire}, but it has a very weak impact on the magnetic proximity effect in this system\cite{tong2019magnetic}.

In such a moiré pattern, the local atomic registry varies periodically in real space, while the local bandgap and spin splitting strength in the BAs layer also vary smoothly and periodically. This can be described by a spin-dependent moiré potential,
\begin{equation}\label{eqi1}
 V_{\tau}(\mathbf{r}) = v_{\tau} \sum_{i=1,2,3}\cos(\mathbf{G}_i\cdot\mathbf{r}  + \varphi_{\tau})  ,
\end{equation}
\noindent where $\mathbf{G}_i$ is the reciprocal lattice vector with $\mathbf{G}_1 = \left[0, {4\pi}/{\sqrt{3}a}\right]$ and $ \mathbf{G}_3 = \hat{C}_3 \mathbf{G}_2 = \hat{C}_3^2 \mathbf{G}_1 $. The mapping from the moiré supercell to the monolayer BAs unit cell is denoted by $\mathbf{r}$, where $\mathbf{r}(\mathbf{R}) = \mathbf{R} - \mathbf{R'} = \mathbf{R} - (1+\delta) \hat{C}_{\theta} \mathbf{R}$. The fitting parameters $v_{\tau}$ and $\varphi_{\tau}$ are involved, where $\tau = \pm 1 $ represents the spin index \cite{yu2017moire,tong2019magnetic}. We have neglected the possibility of the lattice relaxation that may exist under small twist angle\cite{zhai2022ultrafast, magorrian2021multifaceted}, and the effect of the lattice relaxation will be investigated in future work.

Since the local atomic configuration varies periodically in real space, the interlayer distance also varies periodically and can be represented as
\begin{equation}\label{eqi2}
	\ d(\mathbf{r}) = \\
d_0 \sum_{i=1,2,3} \cos(\mathbf{G}_i \cdot \mathbf{r} + \varphi_d)   ,
\end{equation}
\noindent where $d_0$ and $\varphi_d$ are spin-independent fitting parameters. When a perpendicular electric field is applied into the central scattering region, the variation of the interlayer distance can lead to the modulation of the Stark shift energy, which acts as a spin-independent moiré potential, 
\begin{equation}\label{eqi3}
	 V(\mathbf{r}) = e\epsilon d(\mathbf{r}) = E_d\sum_{i=1,2,3} \cos(\mathbf{G}_i \cdot \mathbf{r} + \varphi_d)    ,
\end{equation}
\noindent where $E_d = e\epsilon d_0 $, that $\epsilon$ represents the perpendicular electric field which can be positive or negative.

The tight-binding model can be established for monolayer BAs, where the influence of CrI$_3$ in the scattering region can be represented by applying spin-dependent moiré potential. If a perpendicular electric field is applied, a spin-independent electric field modulation moiré potential is added. Consequently, the Hamiltonian of our research model can be expressed as
\begin{equation}\label{eqi4}
 H = \sum_{i,\tau} (\varepsilon_i + V_{i,\tau} + V_i) c_{i,\tau}^\dagger c_{i,\tau} + t \sum_{<i,j>} c_{i,\tau}^\dagger c_{j,\tau} .
\end{equation}
\noindent Here, $\varepsilon_i$ and $t$ represent the onsite energy and hopping in monolayer BAs, respectively. $V_{i,\tau}$ and $V_i$ represent spin-dependent moiré potential and spin-independent electric field modulation moiré potential onsite $i$, respectively. 

\begin{figure*}
	\centering
	\includegraphics[width=17cm]{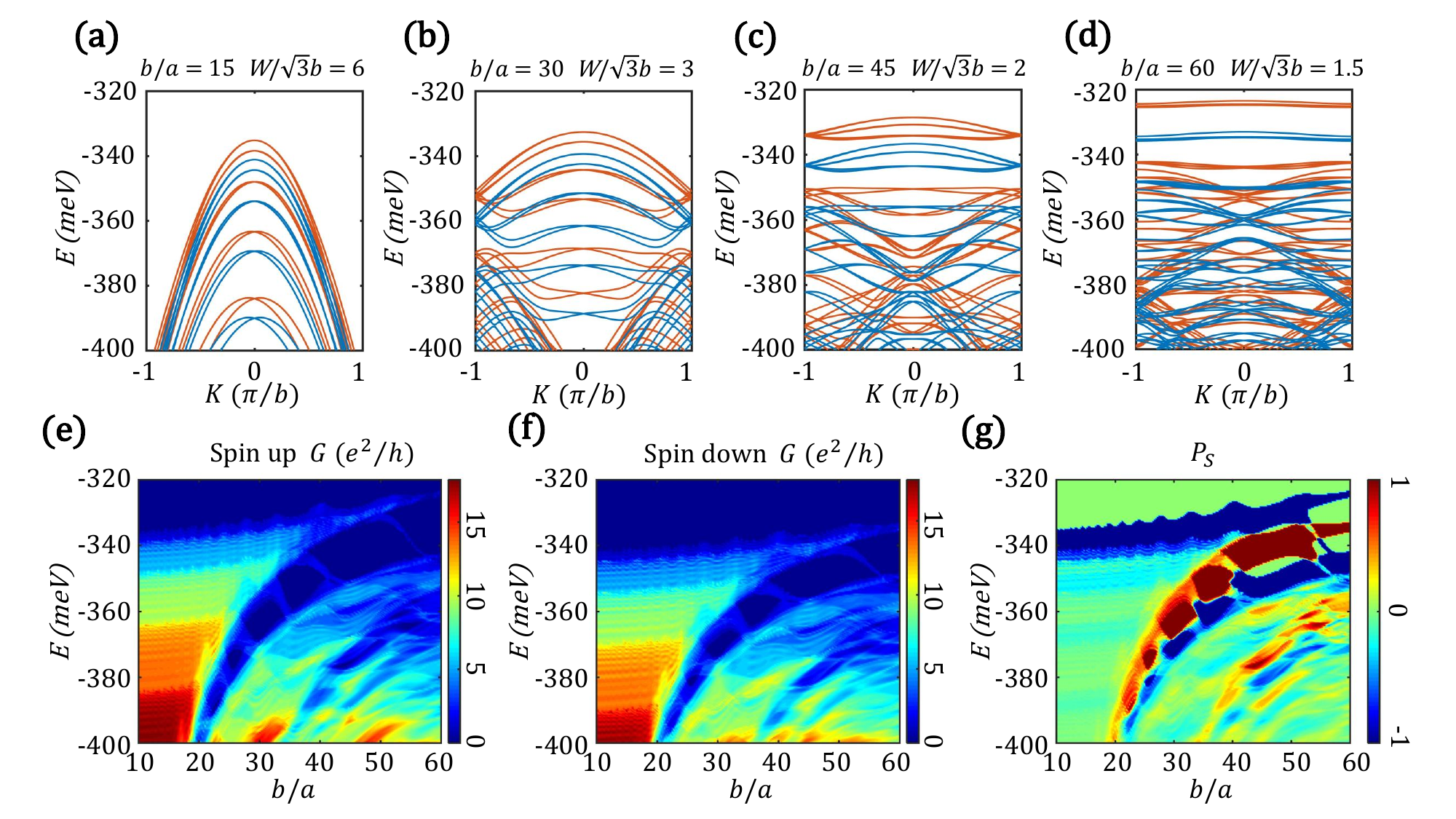}
	\caption{\label{fig:a2}The spin-dependent minband structures at different moiré period $b/a$ and the spin-dependent conductance and spin polarization as functions of incident energy and the moiré period $b/a$. ${W}/\sqrt{3}b$ is the number of periods of the moiré superlattice in the direction of nanoribbon width within the central scattering region, which is closely associated with the number of subbands of the minband. The corresponding twisting angles for the moiré period of $b/a = 15$ and $b/a = 60$ are approximately $\mathrm{\theta \approx 3.75^\circ}$ and $\mathrm{\theta \approx 0.6^\circ}$. 
	}
\end{figure*}

In order to study the quantum transport phenomena of the system, we employed the efficient Non-equilibrium Green's function method \cite{ando1991quantum}. This method has been applied to compute the valley pumping of electrons or holes at non-magnetic disorders \cite{an2017realization}, and it is also very convenient and effective for calculating the spin polarization. The transmission of electrons with incident energy $E$ can be calculated by the formula,
\begin{equation}\label{eqi5}
 T(E) = \text{Tr}[\Gamma_L(E)G(E)\Gamma_R(E)G^\dagger(E)],
\end{equation}
\noindent where $G(E)$ is the Green's function in the central scattering region that satisfies the equation,
\begin{equation}\label{eqi6}
(EI-H-\Sigma_L(E)-\Sigma_R(E))G(E) = I ,
\end{equation}

\noindent where $I$ represents the identity matrix, $H$ is the Hamiltonian of the central scattering region, $\Sigma_L(E)$ and $\Sigma_R(E)$ are the left and right self-energies, respectively, which describe the influence of the semi-infinite monolayer BAs electrodes on the central scattering region from the left and right leads. $\Gamma_L(E)$ and $\Gamma_R(E)$ are the left and right linewidth functions, respectively, and they describe the degree to which the interaction between the electrodes and the central scattering region affects the scattering states. They can be obtained by calculating the imaginary part of the left and right self-energies,

\begin{equation}\label{eqi7}
	 \Gamma_L(E) = -2\text{Im}[\Sigma_L(E)],
\end{equation}
\begin{equation}\label{eqi8}
	\Gamma_R(E) = -2\text{Im}[\Sigma_R(E)].
\end{equation}
According to the Landauer-Büttiker formula, the spin-dependent conductance can be expressed as $G_\tau(E) = \frac{e^2}{h}T_\tau(E)$. The spin polarization is defined as 
\begin{equation}\label{eqi9}
P_S(E)=\frac{G_{-1}(E)-G_{+1}(E)}{G_{-1}(E)+G_{+1}(E)}.
\end{equation}

\section{RESULTS AND DISCUSSION}\label{sec3}

The miniband in the mini-Brillouin zone of the superlattice can be obtained by inwardly folding the Brillouin zone of monolayer BAs \cite{tong2019magnetic}. The intervalley scattering is negligible since the moiré potential is smooth, but the energy gap can be opened due to the intravalley scattering arising from the moiré modulation. The wave function of the band-edge state is localized around the moiré confinement center. With the increase of moiré period, the band-edge state becomes more localized and the overlap of the wave functions between these states decreases. Accordingly, the inter-moiré hopping decreases, the miniband becomes flat and the two spin species are separated \cite{tong2019magnetic}. The transport properties of nanoribbons with periodic superlattices are closely related to these miniband structure. For superlattices with long periods, the spin-dependent minigap makes it possible to obtain spin-polarized transport properties.

The length and width of the central scattering region in the device are signed as ${L}=360a$ and ${W} = 90\sqrt{3}a$, respectively. The moiré period is set to be $b/a=45$, corresponding to a twist angle of $\theta \approx  1.03^\circ$ between the bilayer heterostructure. With this configuration, the scattering region extends over eight periods of the moiré pattern in length and two periods in width. If the moiré period is set to be $b/a=15$, a twist angle of $\theta \approx 3.75^\circ$ follows. The onsite energy parameters in monolayer BAs, namely $\epsilon_B = -\epsilon_{As} = 0.38 eV$, are determined through first-principles calculations. Here, we focus on the valence band edge. The spin-dependent moiré potential parameters are determined as follows: $\varphi_{+1}=152.3^\circ$, $\varphi_{-1}=156^\circ$. The parameters $v_{+1}$ and $v_{-1}$ are used to adjust the range of the moiré potential $V_{+1}(\mathbf{r}) \sim (0,83)  meV$ and $V_{-1}(\mathbf{r}) \sim (0,70)  meV$, respectively. Here, $\varphi_{\tau}$ determines the shape of the moiré potential, while $v_{\tau}$ determines the strength of interlayer coupling. The interlayer atomic registry in the A high-symmetry point determines the weak interlayer interaction here, with negligible magnetic proximity effect and no spin splitting. Therefore, the potential energy here for both spin-up and spin-down is defined as $0 meV$. On the other hand, at the H high-symmetry point, the interlayer interaction and magnetic proximity effect are strongest, resulting in significant spin splitting. The potential energy here for the spin-up is $83 meV$, while for the spin-down is $70 meV$, leading to a spin splitting of approximately $12.7 meV$. These desired moiré potentials can be obtained by adjusting the aforementioned parameters. The spin-independent moiré potential parameters are determined as follows: $d_0=0.12\text{Å}$ and $\varphi_d=167^\circ$. These parameters have been derived through fitting procedures based on first-principles calculations \cite{tong2019magnetic}.

As shown in Fig.~\ref{fig:a1}(c), the middle panel illustrates the spin-dependent conductance as a function of the incident energy $E$ for moiré period with $b/a=15$. Electron localization in the moiré potential is very weak, so the corresponding miniband is not flat and there is no minigap in the conductance. The plateau-like structure of the spin-up and spin-down conductances is observed. This phenomenon can be understood by examining the spin-dependent minband structure at a moiré period of $b/a=15$ depicted in Fig.~\ref{fig:a2}(a). However, the resonance structure is superimposed on the conductances at the transition from one plateau to another \cite{mu2022valley}. These resonances are associated with multiple reflection in the moiré potential. At this twist angle, the magnetic proximity effect is not significant and the spin is not split in energy. The difference between the spin-up and spin-down moiré potentials results in a weaker spin polarization, as shown in the bottom panel of Fig.~\ref{fig:a1}(b).

When the moiré period is adjusted to $b/a=45$, the miniband of the system becomes flat with a sizable minigap, which is on the same order of magnitude as the spin splitting induced by the magnetic proximity effects. The spin-dependent conductance as a function of incident energy is plotted in the top panel of Fig.~\ref{fig:a1}(c). The conductance clearly exhibits miniband transport properties with the presence of spin-up and spin-down minigaps. Furthermore, both of the spin-up and spin-down conductances exhibit similar trends for the incident energy. They exhibit a relative shift along the incident energy that is approximately equal to magnitude of the spin splitting. At this small twist angle, the minigaps for both spin-up and spin-down conductances intercalate into each other, leading to a staggered spin polarization pattern as depicted in the bottom panel of Fig.~\ref{fig:a1}(c). Therefore, the intensity and direction of spin polarization can be controlled by the incident energy.

\begin{figure*}
	\centering
	\includegraphics[width=17cm]{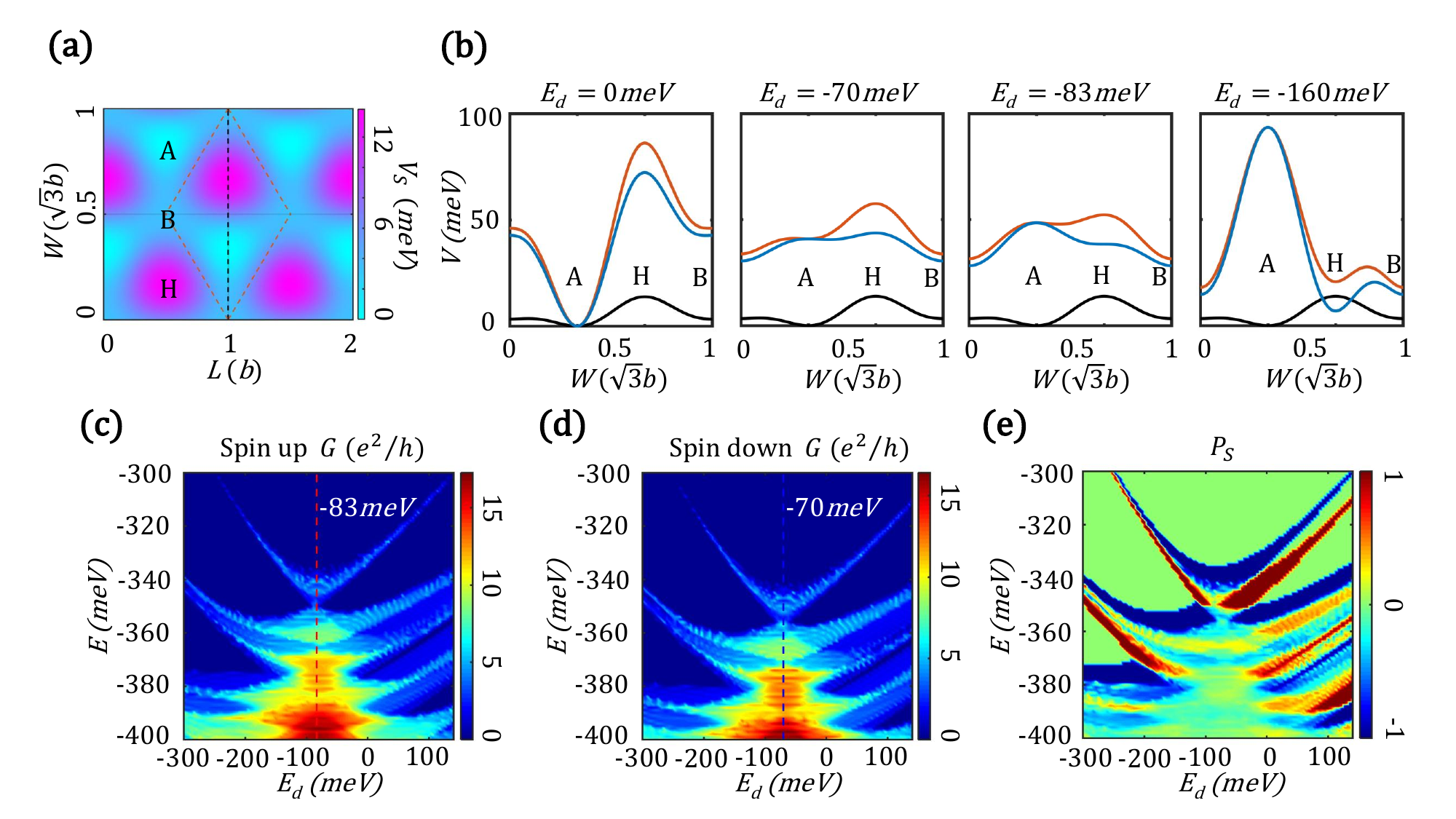}
	\caption{\label{fig:a3}The moiré potential under different electric field modulations and the spin-dependent conductance and spin polarization under various electric field modulations. $E_d$ is the electric field modulation parameter in Eq. (\ref{eqi3}). (a) demonstrated the manifestation of spin splitting in real space where $V_S(\mathbf{r}) = V_{+1}(\mathbf{r}) - V_{-1}(\mathbf{r})$. The moiré potential under different electric field modulations in (b) is given by the black dashed line in (a), with the red and blue lines representing the potential for spin-up and spin-down, respectively. The black line denotes the spin-splitting potential. The moiré period is set as $b/a = 45$. Both the conductance of spin-up and spin-down exhibit a certain symmetry with respect to the electric field modulation parameter $E_d$, where the approximate axis of symmetry for the conductance of spin-up is $E_d = -83 meV$ in (c), and the approximate axis of symmetry for the conductance of spin-down is $E_d = -70 meV$ in (d).
	}
\end{figure*}

The moiré period size has a significant impact on the miniband transport and spin polarization. In the following, we investigated the minibands at different moiré periods. The conductance and spin polarization as functions of incident energy and moiré period, as depicted in Fig.~\ref{fig:a2}. In Figs.~\ref{fig:a2}(a), (b), (c), and (d), as $b/a$ increases, the minibands gradually become flat and narrow, while the minigaps widen. At $b/a=15$, there are no gaps in the energy range of interest, while at $b/a=30$, the minigaps gradually emerge and the minibands become flat. As $b/a$ reaches $45$, it become even flatter, and the spin degeneracy of the first miniband is completely lifted, with the spin-down minband fully penetrating into the spin-up minigap. At $b/a=60$, the bands become even flatter. Furthermore, when the width of the nanoribbon ${W}$ is fixed, the number of subbands in the miniband decreases as the moiré period $b/a$ increases. From a real space perspective, this is due to the decreasing number of superlattice periods along the nanoribbon width direction ${W}/\sqrt{3}b$. This leads to the presence of plateau-like transport in the conductance around $b/a=10$ to $b/a=30$ and the steps gradually disappearing, as shown in Figs.~\ref{fig:a2}(e) and (f).

In the range of moiré period from $b/a=10$ to $b/a=30$, we can see that both of the spin-up and spin-down conductance exhibit plateau-like features. This corresponds to the subbands of the first moiré miniband, as illustrated in Fig.~\ref{fig:a2}(a), and these subbands remain invariant within certain ranges of the moiré period ($b/a$).  Electron incident energy traverses multiple subbands, each corresponding to an integer multiple of the quantized conductance, ${e^2}/{h}$. With the increasing of moiré period which is more than $b/a=40$, the miniband transport properties gradually exhibit. Moreover, the width of the miniband gradually becomes smaller and minigap becomes larger as the moiré period increases. This is because charges are becoming more and more localized at the center of moiré potential. All of these results are consistent with our previous discussions. In addition, due to the presence of spin splitting, the spin-up and spin-down conductances exhibit a relative shift at the moiré period and the incident energy resulting in the remarkable spin polarization. Moreover, when the size of the minigap is modulated to be comparable to the spin splitting, the spin polarization becomes more pronounced. Therefore, setting the moiré period to be around $b/a=45$ is preferable. Here, not only the first and second minibands exhibit 100\% spin polarization, but some of the following minibands also display an staggered spin polarization.

The spin-dependent transport properties of the system with the modulation of an external perpendicular electric field are also investigated. To visually illustrate the impact of electric field modulation on the transport properties, we present the moiré potential under different electric field modulations as shown in Fig.~\ref{fig:a3}(b). A clear observation is the gradual “flattening” of the moiré potential as the electric field modulation parameter varies from approximately $E_d = 0 meV$ to $E_d = -80 meV$. As a result, the localization of electrons weakens, and the minibands gradually widen while the minigaps narrow, as depicted in the conductance plots in Figs.~\ref{fig:a3}(c) and (d). During the process of electric field modulation parameter varies from $E_d = -80 meV$ to $E_d = -160 meV$, the moiré potential becomes nonuniform again, exhibiting a prominent energy potential at the high-symmetry point A. This results in electron localization and increasingly larger bandgaps. In summary, electric field modulation exhibits a certain degree of symmetry in its impact on electron localization, resulting in symmetric effects on miniband structure and conductance. 

Additionally, as shown in Fig.~\ref{fig:a3}(b), the moiré potential for the spin down state is most flattened when $E_d = -70 meV$, while for the spin up state, it becomes most flattened at $E_d = -83 meV$. Consequently, the positions of the symmetry axis for the conductances differ for spin down and spin up, occurring at $E_d = -70 meV$ and $E_d = -83 meV$, respectively, as illustrated in Figs.~\ref{fig:a3}(c) and (d). Not only they have different axis of symmetry, but also have an overall relative shift along the electric field. This is the specific manifestation of the differential response of spin-up and spin-down conductance to the electric field. This results in the symmetric spin polarization form, as shown in Fig.~\ref{fig:a3}(e). In this configuration, we can adjust spin polarization not only by changing the Fermi level in the lead but also by modifying the electron's localization through applying electric field modulation in the scattering region. Moreover, under different twist angles, although the size of the moiré period may vary, the shape of the moiré potential remains consistent. Therefore, this modulation should exist at different twist angles, with the most pronounced manifestation observed around $b/a = 45$.

\begin{figure}
	\centering
	\includegraphics[width=7.5cm]{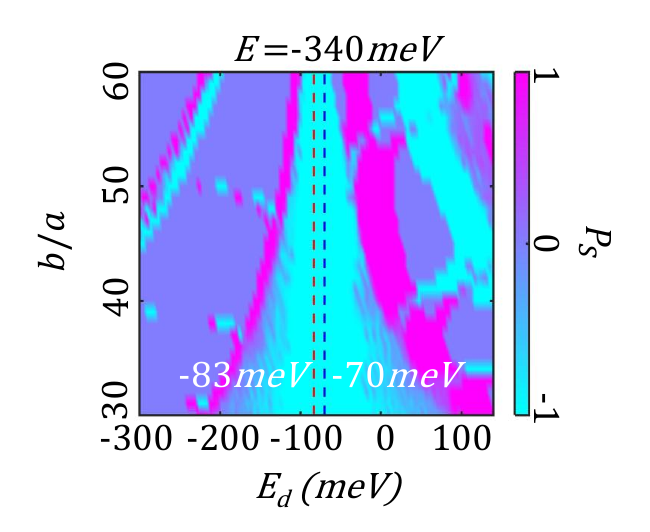}
	\caption{\label{fig:a4} The functional relationship between moiré period $b/a$ and electric field modulation $E_d$ on spin polarization with fixed electron incident energy $E = -340meV$.
	}
\end{figure}

\begin{figure}
	\centering
	\includegraphics[width=9cm]{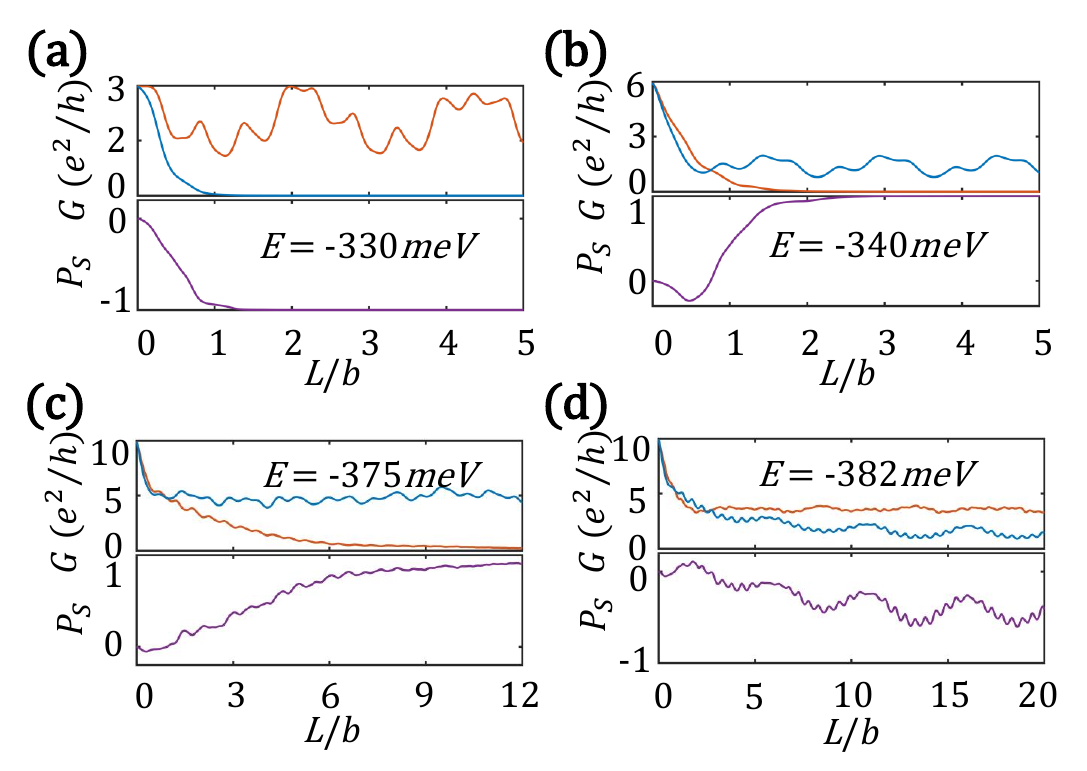}
	\caption{\label{fig:a5} The relationship between the the number of moiré periods in the length direction of the scattering region(${L}/{b}$) and spin resolution conductance and spin polarization at fixed electron incident energies. The electric field is absent and the moiré period is $b/a=45$.
	}
\end{figure}

For further research the effect of electric field modulation on spin polarization, we plot the spin polarization as a function of the electric field and the moiré periods with fixed incident energy as shown in Fig.~\ref{fig:a4}. The symmetric spin polarization phenomenon exists not only in the case of moiré period $b/a = 45$, but also in other moiré periods. This demonstrates that spin polarization reversal modulation can be achieved solely through electric field modulation in the scattering region without altering the Fermi level of the lead.

We also examined the effect of the scattering region length on the spin-dependent conductance and spin polarization. Figures~\ref{fig:a5} show the spin-dependent conductance and spin polarization as functions of the scattering region length ${L}$ for the various incident energies in the spin-resolved minigaps. The spin-dependent conductances exhibit obvious resonance transport properties, and one of the spin-dependent conductances decays to zero when the length of the scattering region increases. When the length of the scattering region is short, the scattering region contains only a few periods, and there is no significant spin polarization because there is only the scattering of the moiré potential without forming the band structure in the scattering region. When the length of the scattering region becomes longer, the scattering region contains a sufficient number of periods, then the spin-dependent energy gap appears in the system and 100\% spin polarization is obtained.  Finite periods are sufficient to achieve high efficiency spin filter. For example, it only takes two or three periods to realize almost 100\% spin polarization at low energy for hole as shown in Figs.~\ref{fig:a5}(a) and (b). More superlattice periods are required in order to realize a remarkable spin polarization in the high energy for hole. For instance, a large spin polarization is obtained remarkably, by using ten to fifteen moiré periods as shown in Figs.~\ref{fig:a5}(c) and (d). This is because the minigap in the lower energy region is formed more easily than that in the high energy region.

\section{Conclusions} \label{sec4}

In summary, we have investigate the spin-polarized transport properties in a moiré superlattices formed by a monolayer BAs stacked on a twisted 2D ferromagnet CrI$_{3}$. We find that the conductance exhibits spin-resolved miniband transport properties at a small twist angle because of the magnetic proximity effect associated with the moiré superlattices of long period. A remarkable spin polarization can be realized when the incident energy lies in the spin-resolved gaps. We show that the direction and strength of the spin polarization can further be tuned by a perpendicular electric field because the interlayer distance of the heterojunction is also moiré modifiable. A practical spin polarization can be generated simply by passing the charge current through a finite moiré period. These results point to an unexpected but exciting opportunity to build spin functionality in magnetic moiré electronics.

\begin{acknowledgments}
We thank Professor Qing-Jun Tong for many helpful discussions.
This work was supported by the National Natural Science Foundation of China (No. 12074096) and Hebei Province Natural Science Foundation of China (No. A2021208013).
J.L. acknowledge support by the National Natural Science Foundation of China (No. 12274305).
\end{acknowledgments}

\bibliography{mymy}

\begin{thebibliography}{37}%
\makeatletter
\providecommand \@ifxundefined [1]{%
 \@ifx{#1\undefined}
}%
\providecommand \@ifnum [1]{%
 \ifnum #1\expandafter \@firstoftwo
 \else \expandafter \@secondoftwo
 \fi
}%
\providecommand \@ifx [1]{%
 \ifx #1\expandafter \@firstoftwo
 \else \expandafter \@secondoftwo
 \fi
}%
\providecommand \natexlab [1]{#1}%
\providecommand \enquote  [1]{``#1''}%
\providecommand \bibnamefont  [1]{#1}%
\providecommand \bibfnamefont [1]{#1}%
\providecommand \citenamefont [1]{#1}%
\providecommand \href@noop [0]{\@secondoftwo}%
\providecommand \href [0]{\begingroup \@sanitize@url \@href}%
\providecommand \@href[1]{\@@startlink{#1}\@@href}%
\providecommand \@@href[1]{\endgroup#1\@@endlink}%
\providecommand \@sanitize@url [0]{\catcode `\\12\catcode `\$12\catcode
  `\&12\catcode `\#12\catcode `\^12\catcode `\_12\catcode `\%12\relax}%
\providecommand \@@startlink[1]{}%
\providecommand \@@endlink[0]{}%
\providecommand \url  [0]{\begingroup\@sanitize@url \@url }%
\providecommand \@url [1]{\endgroup\@href {#1}{\urlprefix }}%
\providecommand \urlprefix  [0]{URL }%
\providecommand \Eprint [0]{\href }%
\providecommand \doibase [0]{https://doi.org/}%
\providecommand \selectlanguage [0]{\@gobble}%
\providecommand \bibinfo  [0]{\@secondoftwo}%
\providecommand \bibfield  [0]{\@secondoftwo}%
\providecommand \translation [1]{[#1]}%
\providecommand \BibitemOpen [0]{}%
\providecommand \bibitemStop [0]{}%
\providecommand \bibitemNoStop [0]{.\EOS\space}%
\providecommand \EOS [0]{\spacefactor3000\relax}%
\providecommand \BibitemShut  [1]{\csname bibitem#1\endcsname}%
\let\auto@bib@innerbib\@empty
\bibitem [{\citenamefont {Cao}\ \emph {et~al.}(2018{\natexlab{a}})\citenamefont
  {Cao}, \citenamefont {Fatemi}, \citenamefont {Fang}, \citenamefont
  {Watanabe}, \citenamefont {Taniguchi}, \citenamefont {Kaxiras},\ and\
  \citenamefont {Jarillo-Herrero}}]{cao2018unconventional}%
  \BibitemOpen
  \bibfield  {author} {\bibinfo {author} {\bibfnamefont {Y.}~\bibnamefont
  {Cao}}, \bibinfo {author} {\bibfnamefont {V.}~\bibnamefont {Fatemi}},
  \bibinfo {author} {\bibfnamefont {S.}~\bibnamefont {Fang}}, \bibinfo {author}
  {\bibfnamefont {K.}~\bibnamefont {Watanabe}}, \bibinfo {author}
  {\bibfnamefont {T.}~\bibnamefont {Taniguchi}}, \bibinfo {author}
  {\bibfnamefont {E.}~\bibnamefont {Kaxiras}},\ and\ \bibinfo {author}
  {\bibfnamefont {P.}~\bibnamefont {Jarillo-Herrero}},\ }\href@noop {}
  {\bibfield  {journal} {\bibinfo  {journal} {Nature}\ }\textbf {\bibinfo
  {volume} {556}},\ \bibinfo {pages} {43} (\bibinfo {year}
  {2018}{\natexlab{a}})}\BibitemShut {NoStop}%
\bibitem [{\citenamefont {Yankowitz}\ \emph {et~al.}(2019)\citenamefont
  {Yankowitz}, \citenamefont {Chen}, \citenamefont {Polshyn}, \citenamefont
  {Zhang}, \citenamefont {Watanabe}, \citenamefont {Taniguchi}, \citenamefont
  {Graf}, \citenamefont {Young},\ and\ \citenamefont
  {Dean}}]{yankowitz2019tuning}%
  \BibitemOpen
  \bibfield  {author} {\bibinfo {author} {\bibfnamefont {M.}~\bibnamefont
  {Yankowitz}}, \bibinfo {author} {\bibfnamefont {S.}~\bibnamefont {Chen}},
  \bibinfo {author} {\bibfnamefont {H.}~\bibnamefont {Polshyn}}, \bibinfo
  {author} {\bibfnamefont {Y.}~\bibnamefont {Zhang}}, \bibinfo {author}
  {\bibfnamefont {K.}~\bibnamefont {Watanabe}}, \bibinfo {author}
  {\bibfnamefont {T.}~\bibnamefont {Taniguchi}}, \bibinfo {author}
  {\bibfnamefont {D.}~\bibnamefont {Graf}}, \bibinfo {author} {\bibfnamefont
  {A.~F.}\ \bibnamefont {Young}},\ and\ \bibinfo {author} {\bibfnamefont
  {C.~R.}\ \bibnamefont {Dean}},\ }\href@noop {} {\bibfield  {journal}
  {\bibinfo  {journal} {Science}\ }\textbf {\bibinfo {volume} {363}},\ \bibinfo
  {pages} {1059} (\bibinfo {year} {2019})}\BibitemShut {NoStop}%
\bibitem [{\citenamefont {Sharpe}\ \emph {et~al.}(2019)\citenamefont {Sharpe},
  \citenamefont {Fox}, \citenamefont {Barnard}, \citenamefont {Finney},
  \citenamefont {Watanabe}, \citenamefont {Taniguchi}, \citenamefont
  {Kastner},\ and\ \citenamefont {Goldhaber-Gordon}}]{sharpe2019emergent}%
  \BibitemOpen
  \bibfield  {author} {\bibinfo {author} {\bibfnamefont {A.~L.}\ \bibnamefont
  {Sharpe}}, \bibinfo {author} {\bibfnamefont {E.~J.}\ \bibnamefont {Fox}},
  \bibinfo {author} {\bibfnamefont {A.~W.}\ \bibnamefont {Barnard}}, \bibinfo
  {author} {\bibfnamefont {J.}~\bibnamefont {Finney}}, \bibinfo {author}
  {\bibfnamefont {K.}~\bibnamefont {Watanabe}}, \bibinfo {author}
  {\bibfnamefont {T.}~\bibnamefont {Taniguchi}}, \bibinfo {author}
  {\bibfnamefont {M.}~\bibnamefont {Kastner}},\ and\ \bibinfo {author}
  {\bibfnamefont {D.}~\bibnamefont {Goldhaber-Gordon}},\ }\href@noop {}
  {\bibfield  {journal} {\bibinfo  {journal} {Science}\ }\textbf {\bibinfo
  {volume} {365}},\ \bibinfo {pages} {605} (\bibinfo {year}
  {2019})}\BibitemShut {NoStop}%
\bibitem [{\citenamefont {Serlin}\ \emph {et~al.}(2020)\citenamefont {Serlin},
  \citenamefont {Tschirhart}, \citenamefont {Polshyn}, \citenamefont {Zhang},
  \citenamefont {Zhu}, \citenamefont {Watanabe}, \citenamefont {Taniguchi},
  \citenamefont {Balents},\ and\ \citenamefont {Young}}]{serlin2020intrinsic}%
  \BibitemOpen
  \bibfield  {author} {\bibinfo {author} {\bibfnamefont {M.}~\bibnamefont
  {Serlin}}, \bibinfo {author} {\bibfnamefont {C.}~\bibnamefont {Tschirhart}},
  \bibinfo {author} {\bibfnamefont {H.}~\bibnamefont {Polshyn}}, \bibinfo
  {author} {\bibfnamefont {Y.}~\bibnamefont {Zhang}}, \bibinfo {author}
  {\bibfnamefont {J.}~\bibnamefont {Zhu}}, \bibinfo {author} {\bibfnamefont
  {K.}~\bibnamefont {Watanabe}}, \bibinfo {author} {\bibfnamefont
  {T.}~\bibnamefont {Taniguchi}}, \bibinfo {author} {\bibfnamefont
  {L.}~\bibnamefont {Balents}},\ and\ \bibinfo {author} {\bibfnamefont
  {A.}~\bibnamefont {Young}},\ }\href@noop {} {\bibfield  {journal} {\bibinfo
  {journal} {Science}\ }\textbf {\bibinfo {volume} {367}},\ \bibinfo {pages}
  {900} (\bibinfo {year} {2020})}\BibitemShut {NoStop}%
\bibitem [{\citenamefont {Cao}\ \emph {et~al.}(2018{\natexlab{b}})\citenamefont
  {Cao}, \citenamefont {Fatemi}, \citenamefont {Demir}, \citenamefont {Fang},
  \citenamefont {Tomarken}, \citenamefont {Luo}, \citenamefont
  {Sanchez-Yamagishi}, \citenamefont {Watanabe}, \citenamefont {Taniguchi},
  \citenamefont {Kaxiras} \emph {et~al.}}]{cao2018correlated}%
  \BibitemOpen
  \bibfield  {author} {\bibinfo {author} {\bibfnamefont {Y.}~\bibnamefont
  {Cao}}, \bibinfo {author} {\bibfnamefont {V.}~\bibnamefont {Fatemi}},
  \bibinfo {author} {\bibfnamefont {A.}~\bibnamefont {Demir}}, \bibinfo
  {author} {\bibfnamefont {S.}~\bibnamefont {Fang}}, \bibinfo {author}
  {\bibfnamefont {S.~L.}\ \bibnamefont {Tomarken}}, \bibinfo {author}
  {\bibfnamefont {J.~Y.}\ \bibnamefont {Luo}}, \bibinfo {author} {\bibfnamefont
  {J.~D.}\ \bibnamefont {Sanchez-Yamagishi}}, \bibinfo {author} {\bibfnamefont
  {K.}~\bibnamefont {Watanabe}}, \bibinfo {author} {\bibfnamefont
  {T.}~\bibnamefont {Taniguchi}}, \bibinfo {author} {\bibfnamefont
  {E.}~\bibnamefont {Kaxiras}}, \emph {et~al.},\ }\href@noop {} {\bibfield
  {journal} {\bibinfo  {journal} {Nature}\ }\textbf {\bibinfo {volume} {556}},\
  \bibinfo {pages} {80} (\bibinfo {year} {2018}{\natexlab{b}})}\BibitemShut
  {NoStop}%
\bibitem [{\citenamefont {Vizner~Stern}\ \emph {et~al.}(2021)\citenamefont
  {Vizner~Stern}, \citenamefont {Waschitz}, \citenamefont {Cao}, \citenamefont
  {Nevo}, \citenamefont {Watanabe}, \citenamefont {Taniguchi}, \citenamefont
  {Sela}, \citenamefont {Urbakh}, \citenamefont {Hod},\ and\ \citenamefont
  {Ben~Shalom}}]{vizner2021interfacial}%
  \BibitemOpen
  \bibfield  {author} {\bibinfo {author} {\bibfnamefont {M.}~\bibnamefont
  {Vizner~Stern}}, \bibinfo {author} {\bibfnamefont {Y.}~\bibnamefont
  {Waschitz}}, \bibinfo {author} {\bibfnamefont {W.}~\bibnamefont {Cao}},
  \bibinfo {author} {\bibfnamefont {I.}~\bibnamefont {Nevo}}, \bibinfo {author}
  {\bibfnamefont {K.}~\bibnamefont {Watanabe}}, \bibinfo {author}
  {\bibfnamefont {T.}~\bibnamefont {Taniguchi}}, \bibinfo {author}
  {\bibfnamefont {E.}~\bibnamefont {Sela}}, \bibinfo {author} {\bibfnamefont
  {M.}~\bibnamefont {Urbakh}}, \bibinfo {author} {\bibfnamefont
  {O.}~\bibnamefont {Hod}},\ and\ \bibinfo {author} {\bibfnamefont
  {M.}~\bibnamefont {Ben~Shalom}},\ }\href@noop {} {\bibfield  {journal}
  {\bibinfo  {journal} {Science}\ }\textbf {\bibinfo {volume} {372}},\ \bibinfo
  {pages} {1462} (\bibinfo {year} {2021})}\BibitemShut {NoStop}%
\bibitem [{\citenamefont {Regan}\ \emph {et~al.}(2020)\citenamefont {Regan},
  \citenamefont {Wang}, \citenamefont {Jin}, \citenamefont {Bakti~Utama},
  \citenamefont {Gao}, \citenamefont {Wei}, \citenamefont {Zhao}, \citenamefont
  {Zhao}, \citenamefont {Zhang}, \citenamefont {Yumigeta} \emph
  {et~al.}}]{regan2020mott}%
  \BibitemOpen
  \bibfield  {author} {\bibinfo {author} {\bibfnamefont {E.~C.}\ \bibnamefont
  {Regan}}, \bibinfo {author} {\bibfnamefont {D.}~\bibnamefont {Wang}},
  \bibinfo {author} {\bibfnamefont {C.}~\bibnamefont {Jin}}, \bibinfo {author}
  {\bibfnamefont {M.~I.}\ \bibnamefont {Bakti~Utama}}, \bibinfo {author}
  {\bibfnamefont {B.}~\bibnamefont {Gao}}, \bibinfo {author} {\bibfnamefont
  {X.}~\bibnamefont {Wei}}, \bibinfo {author} {\bibfnamefont {S.}~\bibnamefont
  {Zhao}}, \bibinfo {author} {\bibfnamefont {W.}~\bibnamefont {Zhao}}, \bibinfo
  {author} {\bibfnamefont {Z.}~\bibnamefont {Zhang}}, \bibinfo {author}
  {\bibfnamefont {K.}~\bibnamefont {Yumigeta}}, \emph {et~al.},\ }\href@noop {}
  {\bibfield  {journal} {\bibinfo  {journal} {Nature}\ }\textbf {\bibinfo
  {volume} {579}},\ \bibinfo {pages} {359} (\bibinfo {year}
  {2020})}\BibitemShut {NoStop}%
\bibitem [{\citenamefont {Tang}\ \emph
  {et~al.}(2020{\natexlab{a}})\citenamefont {Tang}, \citenamefont {Li},
  \citenamefont {Li}, \citenamefont {Xu}, \citenamefont {Liu}, \citenamefont
  {Barmak}, \citenamefont {Watanabe}, \citenamefont {Taniguchi}, \citenamefont
  {MacDonald}, \citenamefont {Shan} \emph {et~al.}}]{tang2020simulation}%
  \BibitemOpen
  \bibfield  {author} {\bibinfo {author} {\bibfnamefont {Y.}~\bibnamefont
  {Tang}}, \bibinfo {author} {\bibfnamefont {L.}~\bibnamefont {Li}}, \bibinfo
  {author} {\bibfnamefont {T.}~\bibnamefont {Li}}, \bibinfo {author}
  {\bibfnamefont {Y.}~\bibnamefont {Xu}}, \bibinfo {author} {\bibfnamefont
  {S.}~\bibnamefont {Liu}}, \bibinfo {author} {\bibfnamefont {K.}~\bibnamefont
  {Barmak}}, \bibinfo {author} {\bibfnamefont {K.}~\bibnamefont {Watanabe}},
  \bibinfo {author} {\bibfnamefont {T.}~\bibnamefont {Taniguchi}}, \bibinfo
  {author} {\bibfnamefont {A.~H.}\ \bibnamefont {MacDonald}}, \bibinfo {author}
  {\bibfnamefont {J.}~\bibnamefont {Shan}}, \emph {et~al.},\ }\href@noop {}
  {\bibfield  {journal} {\bibinfo  {journal} {Nature}\ }\textbf {\bibinfo
  {volume} {579}},\ \bibinfo {pages} {353} (\bibinfo {year}
  {2020}{\natexlab{a}})}\BibitemShut {NoStop}%
\bibitem [{\citenamefont {Xu}\ \emph {et~al.}(2020)\citenamefont {Xu},
  \citenamefont {Liu}, \citenamefont {Rhodes}, \citenamefont {Watanabe},
  \citenamefont {Taniguchi}, \citenamefont {Hone}, \citenamefont {Elser},
  \citenamefont {Mak},\ and\ \citenamefont {Shan}}]{xu2020correlated}%
  \BibitemOpen
  \bibfield  {author} {\bibinfo {author} {\bibfnamefont {Y.}~\bibnamefont
  {Xu}}, \bibinfo {author} {\bibfnamefont {S.}~\bibnamefont {Liu}}, \bibinfo
  {author} {\bibfnamefont {D.~A.}\ \bibnamefont {Rhodes}}, \bibinfo {author}
  {\bibfnamefont {K.}~\bibnamefont {Watanabe}}, \bibinfo {author}
  {\bibfnamefont {T.}~\bibnamefont {Taniguchi}}, \bibinfo {author}
  {\bibfnamefont {J.}~\bibnamefont {Hone}}, \bibinfo {author} {\bibfnamefont
  {V.}~\bibnamefont {Elser}}, \bibinfo {author} {\bibfnamefont {K.~F.}\
  \bibnamefont {Mak}},\ and\ \bibinfo {author} {\bibfnamefont {J.}~\bibnamefont
  {Shan}},\ }\href@noop {} {\bibfield  {journal} {\bibinfo  {journal} {Nature}\
  }\textbf {\bibinfo {volume} {587}},\ \bibinfo {pages} {214} (\bibinfo {year}
  {2020})}\BibitemShut {NoStop}%
\bibitem [{\citenamefont {Tran}\ \emph {et~al.}(2019)\citenamefont {Tran},
  \citenamefont {Moody}, \citenamefont {Wu}, \citenamefont {Lu}, \citenamefont
  {Choi}, \citenamefont {Kim}, \citenamefont {Rai}, \citenamefont {Sanchez},
  \citenamefont {Quan}, \citenamefont {Singh} \emph
  {et~al.}}]{tran2019evidence}%
  \BibitemOpen
  \bibfield  {author} {\bibinfo {author} {\bibfnamefont {K.}~\bibnamefont
  {Tran}}, \bibinfo {author} {\bibfnamefont {G.}~\bibnamefont {Moody}},
  \bibinfo {author} {\bibfnamefont {F.}~\bibnamefont {Wu}}, \bibinfo {author}
  {\bibfnamefont {X.}~\bibnamefont {Lu}}, \bibinfo {author} {\bibfnamefont
  {J.}~\bibnamefont {Choi}}, \bibinfo {author} {\bibfnamefont {K.}~\bibnamefont
  {Kim}}, \bibinfo {author} {\bibfnamefont {A.}~\bibnamefont {Rai}}, \bibinfo
  {author} {\bibfnamefont {D.~A.}\ \bibnamefont {Sanchez}}, \bibinfo {author}
  {\bibfnamefont {J.}~\bibnamefont {Quan}}, \bibinfo {author} {\bibfnamefont
  {A.}~\bibnamefont {Singh}}, \emph {et~al.},\ }\href@noop {} {\bibfield
  {journal} {\bibinfo  {journal} {Nature}\ }\textbf {\bibinfo {volume} {567}},\
  \bibinfo {pages} {71} (\bibinfo {year} {2019})}\BibitemShut {NoStop}%
\bibitem [{\citenamefont {Seyler}\ \emph {et~al.}(2019)\citenamefont {Seyler},
  \citenamefont {Rivera}, \citenamefont {Yu}, \citenamefont {Wilson},
  \citenamefont {Ray}, \citenamefont {Mandrus}, \citenamefont {Yan},
  \citenamefont {Yao},\ and\ \citenamefont {Xu}}]{seyler2019signatures}%
  \BibitemOpen
  \bibfield  {author} {\bibinfo {author} {\bibfnamefont {K.~L.}\ \bibnamefont
  {Seyler}}, \bibinfo {author} {\bibfnamefont {P.}~\bibnamefont {Rivera}},
  \bibinfo {author} {\bibfnamefont {H.}~\bibnamefont {Yu}}, \bibinfo {author}
  {\bibfnamefont {N.~P.}\ \bibnamefont {Wilson}}, \bibinfo {author}
  {\bibfnamefont {E.~L.}\ \bibnamefont {Ray}}, \bibinfo {author} {\bibfnamefont
  {D.~G.}\ \bibnamefont {Mandrus}}, \bibinfo {author} {\bibfnamefont
  {J.}~\bibnamefont {Yan}}, \bibinfo {author} {\bibfnamefont {W.}~\bibnamefont
  {Yao}},\ and\ \bibinfo {author} {\bibfnamefont {X.}~\bibnamefont {Xu}},\
  }\href@noop {} {\bibfield  {journal} {\bibinfo  {journal} {Nature}\ }\textbf
  {\bibinfo {volume} {567}},\ \bibinfo {pages} {66} (\bibinfo {year}
  {2019})}\BibitemShut {NoStop}%
\bibitem [{\citenamefont {Alexeev}\ \emph {et~al.}(2019)\citenamefont
  {Alexeev}, \citenamefont {Ruiz-Tijerina}, \citenamefont {Danovich},
  \citenamefont {Hamer}, \citenamefont {Terry}, \citenamefont {Nayak},
  \citenamefont {Ahn}, \citenamefont {Pak}, \citenamefont {Lee}, \citenamefont
  {Sohn} \emph {et~al.}}]{alexeev2019resonantly}%
  \BibitemOpen
  \bibfield  {author} {\bibinfo {author} {\bibfnamefont {E.~M.}\ \bibnamefont
  {Alexeev}}, \bibinfo {author} {\bibfnamefont {D.~A.}\ \bibnamefont
  {Ruiz-Tijerina}}, \bibinfo {author} {\bibfnamefont {M.}~\bibnamefont
  {Danovich}}, \bibinfo {author} {\bibfnamefont {M.~J.}\ \bibnamefont {Hamer}},
  \bibinfo {author} {\bibfnamefont {D.~J.}\ \bibnamefont {Terry}}, \bibinfo
  {author} {\bibfnamefont {P.~K.}\ \bibnamefont {Nayak}}, \bibinfo {author}
  {\bibfnamefont {S.}~\bibnamefont {Ahn}}, \bibinfo {author} {\bibfnamefont
  {S.}~\bibnamefont {Pak}}, \bibinfo {author} {\bibfnamefont {J.}~\bibnamefont
  {Lee}}, \bibinfo {author} {\bibfnamefont {J.~I.}\ \bibnamefont {Sohn}}, \emph
  {et~al.},\ }\href@noop {} {\bibfield  {journal} {\bibinfo  {journal}
  {Nature}\ }\textbf {\bibinfo {volume} {567}},\ \bibinfo {pages} {81}
  (\bibinfo {year} {2019})}\BibitemShut {NoStop}%
\bibitem [{\citenamefont {Jin}\ \emph {et~al.}(2019)\citenamefont {Jin},
  \citenamefont {Regan}, \citenamefont {Yan}, \citenamefont {Iqbal
  Bakti~Utama}, \citenamefont {Wang}, \citenamefont {Zhao}, \citenamefont
  {Qin}, \citenamefont {Yang}, \citenamefont {Zheng}, \citenamefont {Shi} \emph
  {et~al.}}]{jin2019observation}%
  \BibitemOpen
  \bibfield  {author} {\bibinfo {author} {\bibfnamefont {C.}~\bibnamefont
  {Jin}}, \bibinfo {author} {\bibfnamefont {E.~C.}\ \bibnamefont {Regan}},
  \bibinfo {author} {\bibfnamefont {A.}~\bibnamefont {Yan}}, \bibinfo {author}
  {\bibfnamefont {M.}~\bibnamefont {Iqbal Bakti~Utama}}, \bibinfo {author}
  {\bibfnamefont {D.}~\bibnamefont {Wang}}, \bibinfo {author} {\bibfnamefont
  {S.}~\bibnamefont {Zhao}}, \bibinfo {author} {\bibfnamefont {Y.}~\bibnamefont
  {Qin}}, \bibinfo {author} {\bibfnamefont {S.}~\bibnamefont {Yang}}, \bibinfo
  {author} {\bibfnamefont {Z.}~\bibnamefont {Zheng}}, \bibinfo {author}
  {\bibfnamefont {S.}~\bibnamefont {Shi}}, \emph {et~al.},\ }\href@noop {}
  {\bibfield  {journal} {\bibinfo  {journal} {Nature}\ }\textbf {\bibinfo
  {volume} {567}},\ \bibinfo {pages} {76} (\bibinfo {year} {2019})}\BibitemShut
  {NoStop}%
\bibitem [{\citenamefont {Bistritzer}\ and\ \citenamefont
  {MacDonald}(2011)}]{bistritzer2011moire}%
  \BibitemOpen
  \bibfield  {author} {\bibinfo {author} {\bibfnamefont {R.}~\bibnamefont
  {Bistritzer}}\ and\ \bibinfo {author} {\bibfnamefont {A.~H.}\ \bibnamefont
  {MacDonald}},\ }\href@noop {} {\bibfield  {journal} {\bibinfo  {journal}
  {Proc. Natl. Acad. Sci.}\ }\textbf {\bibinfo {volume} {108}},\ \bibinfo
  {pages} {12233} (\bibinfo {year} {2011})}\BibitemShut {NoStop}%
\bibitem [{\citenamefont {Yu}\ \emph {et~al.}(2017)\citenamefont {Yu},
  \citenamefont {Liu}, \citenamefont {Tang}, \citenamefont {Xu},\ and\
  \citenamefont {Yao}}]{yu2017moire}%
  \BibitemOpen
  \bibfield  {author} {\bibinfo {author} {\bibfnamefont {H.}~\bibnamefont
  {Yu}}, \bibinfo {author} {\bibfnamefont {G.-B.}\ \bibnamefont {Liu}},
  \bibinfo {author} {\bibfnamefont {J.}~\bibnamefont {Tang}}, \bibinfo {author}
  {\bibfnamefont {X.}~\bibnamefont {Xu}},\ and\ \bibinfo {author}
  {\bibfnamefont {W.}~\bibnamefont {Yao}},\ }\href@noop {} {\bibfield
  {journal} {\bibinfo  {journal} {Sci. Adv.}\ }\textbf {\bibinfo {volume}
  {3}},\ \bibinfo {pages} {e1701696} (\bibinfo {year} {2017})}\BibitemShut
  {NoStop}%
\bibitem [{\citenamefont {Tong}\ \emph {et~al.}(2019)\citenamefont {Tong},
  \citenamefont {Chen},\ and\ \citenamefont {Yao}}]{tong2019magnetic}%
  \BibitemOpen
  \bibfield  {author} {\bibinfo {author} {\bibfnamefont {Q.}~\bibnamefont
  {Tong}}, \bibinfo {author} {\bibfnamefont {M.}~\bibnamefont {Chen}},\ and\
  \bibinfo {author} {\bibfnamefont {W.}~\bibnamefont {Yao}},\ }\href@noop {}
  {\bibfield  {journal} {\bibinfo  {journal} {Phys. Rev. Appl.}\ }\textbf
  {\bibinfo {volume} {12}},\ \bibinfo {pages} {024031} (\bibinfo {year}
  {2019})}\BibitemShut {NoStop}%
\bibitem [{\citenamefont {Wu}\ \emph {et~al.}(2019)\citenamefont {Wu},
  \citenamefont {Lovorn}, \citenamefont {Tutuc}, \citenamefont {Martin},\ and\
  \citenamefont {MacDonald}}]{wu2019topological}%
  \BibitemOpen
  \bibfield  {author} {\bibinfo {author} {\bibfnamefont {F.}~\bibnamefont
  {Wu}}, \bibinfo {author} {\bibfnamefont {T.}~\bibnamefont {Lovorn}}, \bibinfo
  {author} {\bibfnamefont {E.}~\bibnamefont {Tutuc}}, \bibinfo {author}
  {\bibfnamefont {I.}~\bibnamefont {Martin}},\ and\ \bibinfo {author}
  {\bibfnamefont {A.}~\bibnamefont {MacDonald}},\ }\href@noop {} {\bibfield
  {journal} {\bibinfo  {journal} {Phys. Rev. Lett.}\ }\textbf {\bibinfo
  {volume} {122}},\ \bibinfo {pages} {086402} (\bibinfo {year}
  {2019})}\BibitemShut {NoStop}%
\bibitem [{\citenamefont {Yu}\ \emph {et~al.}(2020)\citenamefont {Yu},
  \citenamefont {Chen},\ and\ \citenamefont {Yao}}]{yu2020giant}%
  \BibitemOpen
  \bibfield  {author} {\bibinfo {author} {\bibfnamefont {H.}~\bibnamefont
  {Yu}}, \bibinfo {author} {\bibfnamefont {M.}~\bibnamefont {Chen}},\ and\
  \bibinfo {author} {\bibfnamefont {W.}~\bibnamefont {Yao}},\ }\href@noop {}
  {\bibfield  {journal} {\bibinfo  {journal} {Nat. Sci. Rev.}\ }\textbf
  {\bibinfo {volume} {7}},\ \bibinfo {pages} {12} (\bibinfo {year}
  {2020})}\BibitemShut {NoStop}%
\bibitem [{\citenamefont {Zhai}\ and\ \citenamefont
  {Yao}(2022)}]{zhai2022ultrafast}%
  \BibitemOpen
  \bibfield  {author} {\bibinfo {author} {\bibfnamefont {D.}~\bibnamefont
  {Zhai}}\ and\ \bibinfo {author} {\bibfnamefont {W.}~\bibnamefont {Yao}},\
  }\href@noop {} {\bibfield  {journal} {\bibinfo  {journal} {Nat. Sci.}\
  }\textbf {\bibinfo {volume} {2}},\ \bibinfo {pages} {e20210101} (\bibinfo
  {year} {2022})}\BibitemShut {NoStop}%
\bibitem [{\citenamefont {Tong}\ \emph {et~al.}(2017)\citenamefont {Tong},
  \citenamefont {Yu}, \citenamefont {Zhu}, \citenamefont {Wang}, \citenamefont
  {Xu},\ and\ \citenamefont {Yao}}]{tong2017topological}%
  \BibitemOpen
  \bibfield  {author} {\bibinfo {author} {\bibfnamefont {Q.}~\bibnamefont
  {Tong}}, \bibinfo {author} {\bibfnamefont {H.}~\bibnamefont {Yu}}, \bibinfo
  {author} {\bibfnamefont {Q.}~\bibnamefont {Zhu}}, \bibinfo {author}
  {\bibfnamefont {Y.}~\bibnamefont {Wang}}, \bibinfo {author} {\bibfnamefont
  {X.}~\bibnamefont {Xu}},\ and\ \bibinfo {author} {\bibfnamefont
  {W.}~\bibnamefont {Yao}},\ }\href@noop {} {\bibfield  {journal} {\bibinfo
  {journal} {Nat. Phys.}\ }\textbf {\bibinfo {volume} {13}},\ \bibinfo {pages}
  {356} (\bibinfo {year} {2017})}\BibitemShut {NoStop}%
\bibitem [{\citenamefont {Wu}\ \emph {et~al.}(2018)\citenamefont {Wu},
  \citenamefont {Lovorn}, \citenamefont {Tutuc},\ and\ \citenamefont
  {MacDonald}}]{wu2018hubbard}%
  \BibitemOpen
  \bibfield  {author} {\bibinfo {author} {\bibfnamefont {F.}~\bibnamefont
  {Wu}}, \bibinfo {author} {\bibfnamefont {T.}~\bibnamefont {Lovorn}}, \bibinfo
  {author} {\bibfnamefont {E.}~\bibnamefont {Tutuc}},\ and\ \bibinfo {author}
  {\bibfnamefont {A.~H.}\ \bibnamefont {MacDonald}},\ }\href@noop {} {\bibfield
   {journal} {\bibinfo  {journal} {Phys. Rev. Lett.}\ }\textbf {\bibinfo
  {volume} {121}},\ \bibinfo {pages} {026402} (\bibinfo {year}
  {2018})}\BibitemShut {NoStop}%
\bibitem [{\citenamefont {Yu}\ and\ \citenamefont
  {Zhou}(2023)}]{yu2022phonons}%
  \BibitemOpen
  \bibfield  {author} {\bibinfo {author} {\bibfnamefont {H.}~\bibnamefont
  {Yu}}\ and\ \bibinfo {author} {\bibfnamefont {J.}~\bibnamefont {Zhou}},\
  }\href@noop {} {\bibfield  {journal} {\bibinfo  {journal} {Nat. Sci.}\
  }\textbf {\bibinfo {volume} {3}},\ \bibinfo {pages} {e20220065} (\bibinfo
  {year} {2023})}\BibitemShut {NoStop}%
\bibitem [{\citenamefont {Huang}\ \emph {et~al.}(2017)\citenamefont {Huang},
  \citenamefont {Clark}, \citenamefont {Navarro-Moratalla}, \citenamefont
  {Klein}, \citenamefont {Cheng}, \citenamefont {Seyler}, \citenamefont
  {Zhong}, \citenamefont {Schmidgall}, \citenamefont {McGuire}, \citenamefont
  {Cobden} \emph {et~al.}}]{huang2017layer}%
  \BibitemOpen
  \bibfield  {author} {\bibinfo {author} {\bibfnamefont {B.}~\bibnamefont
  {Huang}}, \bibinfo {author} {\bibfnamefont {G.}~\bibnamefont {Clark}},
  \bibinfo {author} {\bibfnamefont {E.}~\bibnamefont {Navarro-Moratalla}},
  \bibinfo {author} {\bibfnamefont {D.~R.}\ \bibnamefont {Klein}}, \bibinfo
  {author} {\bibfnamefont {R.}~\bibnamefont {Cheng}}, \bibinfo {author}
  {\bibfnamefont {K.~L.}\ \bibnamefont {Seyler}}, \bibinfo {author}
  {\bibfnamefont {D.}~\bibnamefont {Zhong}}, \bibinfo {author} {\bibfnamefont
  {E.}~\bibnamefont {Schmidgall}}, \bibinfo {author} {\bibfnamefont {M.~A.}\
  \bibnamefont {McGuire}}, \bibinfo {author} {\bibfnamefont {D.~H.}\
  \bibnamefont {Cobden}}, \emph {et~al.},\ }\href@noop {} {\bibfield  {journal}
  {\bibinfo  {journal} {Nature}\ }\textbf {\bibinfo {volume} {546}},\ \bibinfo
  {pages} {270} (\bibinfo {year} {2017})}\BibitemShut {NoStop}%
\bibitem [{\citenamefont {Wang}\ \emph {et~al.}(2020)\citenamefont {Wang},
  \citenamefont {Gao}, \citenamefont {Lv}, \citenamefont {Xu},\ and\
  \citenamefont {Xiao}}]{wang2020stacking}%
  \BibitemOpen
  \bibfield  {author} {\bibinfo {author} {\bibfnamefont {C.}~\bibnamefont
  {Wang}}, \bibinfo {author} {\bibfnamefont {Y.}~\bibnamefont {Gao}}, \bibinfo
  {author} {\bibfnamefont {H.}~\bibnamefont {Lv}}, \bibinfo {author}
  {\bibfnamefont {X.}~\bibnamefont {Xu}},\ and\ \bibinfo {author}
  {\bibfnamefont {D.}~\bibnamefont {Xiao}},\ }\href@noop {} {\bibfield
  {journal} {\bibinfo  {journal} {Phys. Rev. Lett.}\ }\textbf {\bibinfo
  {volume} {125}},\ \bibinfo {pages} {247201} (\bibinfo {year}
  {2020})}\BibitemShut {NoStop}%
\bibitem [{\citenamefont {Song}\ \emph {et~al.}(2021)\citenamefont {Song},
  \citenamefont {Sun}, \citenamefont {Anderson}, \citenamefont {Wang},
  \citenamefont {Qian}, \citenamefont {Taniguchi}, \citenamefont {Watanabe},
  \citenamefont {McGuire}, \citenamefont {St{\"o}hr}, \citenamefont {Xiao}
  \emph {et~al.}}]{song2021direct}%
  \BibitemOpen
  \bibfield  {author} {\bibinfo {author} {\bibfnamefont {T.}~\bibnamefont
  {Song}}, \bibinfo {author} {\bibfnamefont {Q.-C.}\ \bibnamefont {Sun}},
  \bibinfo {author} {\bibfnamefont {E.}~\bibnamefont {Anderson}}, \bibinfo
  {author} {\bibfnamefont {C.}~\bibnamefont {Wang}}, \bibinfo {author}
  {\bibfnamefont {J.}~\bibnamefont {Qian}}, \bibinfo {author} {\bibfnamefont
  {T.}~\bibnamefont {Taniguchi}}, \bibinfo {author} {\bibfnamefont
  {K.}~\bibnamefont {Watanabe}}, \bibinfo {author} {\bibfnamefont {M.~A.}\
  \bibnamefont {McGuire}}, \bibinfo {author} {\bibfnamefont {R.}~\bibnamefont
  {St{\"o}hr}}, \bibinfo {author} {\bibfnamefont {D.}~\bibnamefont {Xiao}},
  \emph {et~al.},\ }\href@noop {} {\bibfield  {journal} {\bibinfo  {journal}
  {Science}\ }\textbf {\bibinfo {volume} {374}},\ \bibinfo {pages} {1140}
  (\bibinfo {year} {2021})}\BibitemShut {NoStop}%
\bibitem [{\citenamefont {Sivadas}\ \emph {et~al.}(2018)\citenamefont
  {Sivadas}, \citenamefont {Okamoto}, \citenamefont {Xu}, \citenamefont
  {Fennie},\ and\ \citenamefont {Xiao}}]{sivadas2018stacking}%
  \BibitemOpen
  \bibfield  {author} {\bibinfo {author} {\bibfnamefont {N.}~\bibnamefont
  {Sivadas}}, \bibinfo {author} {\bibfnamefont {S.}~\bibnamefont {Okamoto}},
  \bibinfo {author} {\bibfnamefont {X.}~\bibnamefont {Xu}}, \bibinfo {author}
  {\bibfnamefont {C.~J.}\ \bibnamefont {Fennie}},\ and\ \bibinfo {author}
  {\bibfnamefont {D.}~\bibnamefont {Xiao}},\ }\href@noop {} {\bibfield
  {journal} {\bibinfo  {journal} {Nano lett.}\ }\textbf {\bibinfo {volume}
  {18}},\ \bibinfo {pages} {7658} (\bibinfo {year} {2018})}\BibitemShut
  {NoStop}%
\bibitem [{\citenamefont {Akram}\ \emph {et~al.}(2021)\citenamefont {Akram},
  \citenamefont {LaBollita}, \citenamefont {Dey}, \citenamefont {Kapeghian},
  \citenamefont {Erten},\ and\ \citenamefont {Botana}}]{akram2021moire}%
  \BibitemOpen
  \bibfield  {author} {\bibinfo {author} {\bibfnamefont {M.}~\bibnamefont
  {Akram}}, \bibinfo {author} {\bibfnamefont {H.}~\bibnamefont {LaBollita}},
  \bibinfo {author} {\bibfnamefont {D.}~\bibnamefont {Dey}}, \bibinfo {author}
  {\bibfnamefont {J.}~\bibnamefont {Kapeghian}}, \bibinfo {author}
  {\bibfnamefont {O.}~\bibnamefont {Erten}},\ and\ \bibinfo {author}
  {\bibfnamefont {A.~S.}\ \bibnamefont {Botana}},\ }\href@noop {} {\bibfield
  {journal} {\bibinfo  {journal} {Nano lett.}\ }\textbf {\bibinfo {volume}
  {21}},\ \bibinfo {pages} {6633} (\bibinfo {year} {2021})}\BibitemShut
  {NoStop}%
\bibitem [{\citenamefont {Seyler}\ \emph {et~al.}(2018)\citenamefont {Seyler},
  \citenamefont {Zhong}, \citenamefont {Huang}, \citenamefont {Linpeng},
  \citenamefont {Wilson}, \citenamefont {Taniguchi}, \citenamefont {Watanabe},
  \citenamefont {Yao}, \citenamefont {Xiao}, \citenamefont {McGuire} \emph
  {et~al.}}]{seyler2018valley}%
  \BibitemOpen
  \bibfield  {author} {\bibinfo {author} {\bibfnamefont {K.~L.}\ \bibnamefont
  {Seyler}}, \bibinfo {author} {\bibfnamefont {D.}~\bibnamefont {Zhong}},
  \bibinfo {author} {\bibfnamefont {B.}~\bibnamefont {Huang}}, \bibinfo
  {author} {\bibfnamefont {X.}~\bibnamefont {Linpeng}}, \bibinfo {author}
  {\bibfnamefont {N.~P.}\ \bibnamefont {Wilson}}, \bibinfo {author}
  {\bibfnamefont {T.}~\bibnamefont {Taniguchi}}, \bibinfo {author}
  {\bibfnamefont {K.}~\bibnamefont {Watanabe}}, \bibinfo {author}
  {\bibfnamefont {W.}~\bibnamefont {Yao}}, \bibinfo {author} {\bibfnamefont
  {D.}~\bibnamefont {Xiao}}, \bibinfo {author} {\bibfnamefont {M.~A.}\
  \bibnamefont {McGuire}}, \emph {et~al.},\ }\href@noop {} {\bibfield
  {journal} {\bibinfo  {journal} {Nano Lett.}\ }\textbf {\bibinfo {volume}
  {18}},\ \bibinfo {pages} {3823} (\bibinfo {year} {2018})}\BibitemShut
  {NoStop}%
\bibitem [{\citenamefont {Voroshnin}\ \emph {et~al.}(2022)\citenamefont
  {Voroshnin}, \citenamefont {Tarasov}, \citenamefont {Bokai}, \citenamefont
  {Chikina}, \citenamefont {Senkovskiy}, \citenamefont {Ehlen}, \citenamefont
  {Usachov}, \citenamefont {Gruneis}, \citenamefont {Krivenkov}, \citenamefont
  {Sanchez-Barriga} \emph {et~al.}}]{voroshnin2022direct}%
  \BibitemOpen
  \bibfield  {author} {\bibinfo {author} {\bibfnamefont {V.}~\bibnamefont
  {Voroshnin}}, \bibinfo {author} {\bibfnamefont {A.~V.}\ \bibnamefont
  {Tarasov}}, \bibinfo {author} {\bibfnamefont {K.~A.}\ \bibnamefont {Bokai}},
  \bibinfo {author} {\bibfnamefont {A.}~\bibnamefont {Chikina}}, \bibinfo
  {author} {\bibfnamefont {B.~V.}\ \bibnamefont {Senkovskiy}}, \bibinfo
  {author} {\bibfnamefont {N.}~\bibnamefont {Ehlen}}, \bibinfo {author}
  {\bibfnamefont {D.~Y.}\ \bibnamefont {Usachov}}, \bibinfo {author}
  {\bibfnamefont {A.}~\bibnamefont {Gruneis}}, \bibinfo {author} {\bibfnamefont
  {M.}~\bibnamefont {Krivenkov}}, \bibinfo {author} {\bibfnamefont
  {J.}~\bibnamefont {Sanchez-Barriga}}, \emph {et~al.},\ }\href@noop {}
  {\bibfield  {journal} {\bibinfo  {journal} {ACS nano}\ }\textbf {\bibinfo
  {volume} {16}},\ \bibinfo {pages} {7448} (\bibinfo {year}
  {2022})}\BibitemShut {NoStop}%
\bibitem [{\citenamefont {Norden}\ \emph {et~al.}(2019)\citenamefont {Norden},
  \citenamefont {Zhao}, \citenamefont {Zhang}, \citenamefont {Sabirianov},
  \citenamefont {Petrou},\ and\ \citenamefont {Zeng}}]{norden2019giant}%
  \BibitemOpen
  \bibfield  {author} {\bibinfo {author} {\bibfnamefont {T.}~\bibnamefont
  {Norden}}, \bibinfo {author} {\bibfnamefont {C.}~\bibnamefont {Zhao}},
  \bibinfo {author} {\bibfnamefont {P.}~\bibnamefont {Zhang}}, \bibinfo
  {author} {\bibfnamefont {R.}~\bibnamefont {Sabirianov}}, \bibinfo {author}
  {\bibfnamefont {A.}~\bibnamefont {Petrou}},\ and\ \bibinfo {author}
  {\bibfnamefont {H.}~\bibnamefont {Zeng}},\ }\href@noop {} {\bibfield
  {journal} {\bibinfo  {journal} {Nat. Commun.}\ }\textbf {\bibinfo {volume}
  {10}},\ \bibinfo {pages} {4163} (\bibinfo {year} {2019})}\BibitemShut
  {NoStop}%
\bibitem [{\citenamefont {Tang}\ \emph
  {et~al.}(2020{\natexlab{b}})\citenamefont {Tang}, \citenamefont {Zhang},
  \citenamefont {Lai}, \citenamefont {Tan},\ and\ \citenamefont
  {Gao}}]{tang2020magnetic}%
  \BibitemOpen
  \bibfield  {author} {\bibinfo {author} {\bibfnamefont {C.}~\bibnamefont
  {Tang}}, \bibinfo {author} {\bibfnamefont {Z.}~\bibnamefont {Zhang}},
  \bibinfo {author} {\bibfnamefont {S.}~\bibnamefont {Lai}}, \bibinfo {author}
  {\bibfnamefont {Q.}~\bibnamefont {Tan}},\ and\ \bibinfo {author}
  {\bibfnamefont {W.-b.}\ \bibnamefont {Gao}},\ }\href@noop {} {\bibfield
  {journal} {\bibinfo  {journal} {Adv. Mater.}\ }\textbf {\bibinfo {volume}
  {32}},\ \bibinfo {pages} {1908498} (\bibinfo {year}
  {2020}{\natexlab{b}})}\BibitemShut {NoStop}%
\bibitem [{\citenamefont {Zhong}\ \emph {et~al.}(2020)\citenamefont {Zhong},
  \citenamefont {Seyler}, \citenamefont {Linpeng}, \citenamefont {Wilson},
  \citenamefont {Taniguchi}, \citenamefont {Watanabe}, \citenamefont {McGuire},
  \citenamefont {Fu}, \citenamefont {Xiao}, \citenamefont {Yao} \emph
  {et~al.}}]{zhong2020layer}%
  \BibitemOpen
  \bibfield  {author} {\bibinfo {author} {\bibfnamefont {D.}~\bibnamefont
  {Zhong}}, \bibinfo {author} {\bibfnamefont {K.~L.}\ \bibnamefont {Seyler}},
  \bibinfo {author} {\bibfnamefont {X.}~\bibnamefont {Linpeng}}, \bibinfo
  {author} {\bibfnamefont {N.~P.}\ \bibnamefont {Wilson}}, \bibinfo {author}
  {\bibfnamefont {T.}~\bibnamefont {Taniguchi}}, \bibinfo {author}
  {\bibfnamefont {K.}~\bibnamefont {Watanabe}}, \bibinfo {author}
  {\bibfnamefont {M.~A.}\ \bibnamefont {McGuire}}, \bibinfo {author}
  {\bibfnamefont {K.-M.~C.}\ \bibnamefont {Fu}}, \bibinfo {author}
  {\bibfnamefont {D.}~\bibnamefont {Xiao}}, \bibinfo {author} {\bibfnamefont
  {W.}~\bibnamefont {Yao}}, \emph {et~al.},\ }\href@noop {} {\bibfield
  {journal} {\bibinfo  {journal} {Nat. Nanotechnol.}\ }\textbf {\bibinfo
  {volume} {15}},\ \bibinfo {pages} {187} (\bibinfo {year} {2020})}\BibitemShut
  {NoStop}%
\bibitem [{\citenamefont {Zhao}\ \emph {et~al.}(2023)\citenamefont {Zhao},
  \citenamefont {Ngaloy}, \citenamefont {Ghosh}, \citenamefont {Ershadrad},
  \citenamefont {Gupta}, \citenamefont {Ali}, \citenamefont {Hoque},
  \citenamefont {Karpiak}, \citenamefont {Khokhriakov}, \citenamefont {Polley}
  \emph {et~al.}}]{zhao2023room}%
  \BibitemOpen
  \bibfield  {author} {\bibinfo {author} {\bibfnamefont {B.}~\bibnamefont
  {Zhao}}, \bibinfo {author} {\bibfnamefont {R.}~\bibnamefont {Ngaloy}},
  \bibinfo {author} {\bibfnamefont {S.}~\bibnamefont {Ghosh}}, \bibinfo
  {author} {\bibfnamefont {S.}~\bibnamefont {Ershadrad}}, \bibinfo {author}
  {\bibfnamefont {R.}~\bibnamefont {Gupta}}, \bibinfo {author} {\bibfnamefont
  {K.}~\bibnamefont {Ali}}, \bibinfo {author} {\bibfnamefont {A.~M.}\
  \bibnamefont {Hoque}}, \bibinfo {author} {\bibfnamefont {B.}~\bibnamefont
  {Karpiak}}, \bibinfo {author} {\bibfnamefont {D.}~\bibnamefont
  {Khokhriakov}}, \bibinfo {author} {\bibfnamefont {C.}~\bibnamefont {Polley}},
  \emph {et~al.},\ }\href@noop {} {\bibfield  {journal} {\bibinfo  {journal}
  {Adv. Mater.}\ }\textbf {\bibinfo {volume} {35}},\ \bibinfo {pages} {2209113}
  (\bibinfo {year} {2023})}\BibitemShut {NoStop}%
\bibitem [{\citenamefont {Magorrian}\ \emph {et~al.}(2021)\citenamefont
  {Magorrian}, \citenamefont {Enaldiev}, \citenamefont {Z{\'o}lyomi},
  \citenamefont {Ferreira}, \citenamefont {Fal'ko},\ and\ \citenamefont
  {Ruiz-Tijerina}}]{magorrian2021multifaceted}%
  \BibitemOpen
  \bibfield  {author} {\bibinfo {author} {\bibfnamefont {S.~J.}\ \bibnamefont
  {Magorrian}}, \bibinfo {author} {\bibfnamefont {V.}~\bibnamefont {Enaldiev}},
  \bibinfo {author} {\bibfnamefont {V.}~\bibnamefont {Z{\'o}lyomi}}, \bibinfo
  {author} {\bibfnamefont {F.}~\bibnamefont {Ferreira}}, \bibinfo {author}
  {\bibfnamefont {V.~I.}\ \bibnamefont {Fal'ko}},\ and\ \bibinfo {author}
  {\bibfnamefont {D.~A.}\ \bibnamefont {Ruiz-Tijerina}},\ }\href@noop {}
  {\bibfield  {journal} {\bibinfo  {journal} {Phy. Rev. B}\ }\textbf {\bibinfo
  {volume} {104}},\ \bibinfo {pages} {125440} (\bibinfo {year}
  {2021})}\BibitemShut {NoStop}%
\bibitem [{\citenamefont {Ando}(1991)}]{ando1991quantum}%
  \BibitemOpen
  \bibfield  {author} {\bibinfo {author} {\bibfnamefont {T.}~\bibnamefont
  {Ando}},\ }\href@noop {} {\bibfield  {journal} {\bibinfo  {journal} {Phys.
  Rev. B}\ }\textbf {\bibinfo {volume} {44}},\ \bibinfo {pages} {8017}
  (\bibinfo {year} {1991})}\BibitemShut {NoStop}%
\bibitem [{\citenamefont {An}\ \emph {et~al.}(2017)\citenamefont {An},
  \citenamefont {Xiao}, \citenamefont {Tu}, \citenamefont {Yu}, \citenamefont
  {Fal’ko},\ and\ \citenamefont {Yao}}]{an2017realization}%
  \BibitemOpen
  \bibfield  {author} {\bibinfo {author} {\bibfnamefont {X.-T.}\ \bibnamefont
  {An}}, \bibinfo {author} {\bibfnamefont {J.}~\bibnamefont {Xiao}}, \bibinfo
  {author} {\bibfnamefont {M.-Y.}\ \bibnamefont {Tu}}, \bibinfo {author}
  {\bibfnamefont {H.}~\bibnamefont {Yu}}, \bibinfo {author} {\bibfnamefont
  {V.~I.}\ \bibnamefont {Fal’ko}},\ and\ \bibinfo {author} {\bibfnamefont
  {W.}~\bibnamefont {Yao}},\ }\href@noop {} {\bibfield  {journal} {\bibinfo
  {journal} {Phys. Rev. Lett.}\ }\textbf {\bibinfo {volume} {118}},\ \bibinfo
  {pages} {096602} (\bibinfo {year} {2017})}\BibitemShut {NoStop}%
\bibitem [{\citenamefont {Mu}\ \emph {et~al.}(2022)\citenamefont {Mu},
  \citenamefont {Wang}, \citenamefont {Du}, \citenamefont {An},\ and\
  \citenamefont {Liu}}]{mu2022valley}%
  \BibitemOpen
  \bibfield  {author} {\bibinfo {author} {\bibfnamefont {H.-Y.}\ \bibnamefont
  {Mu}}, \bibinfo {author} {\bibfnamefont {N.-W.}\ \bibnamefont {Wang}},
  \bibinfo {author} {\bibfnamefont {Y.-N.}\ \bibnamefont {Du}}, \bibinfo
  {author} {\bibfnamefont {X.-T.}\ \bibnamefont {An}},\ and\ \bibinfo {author}
  {\bibfnamefont {J.-J.}\ \bibnamefont {Liu}},\ }\href@noop {} {\bibfield
  {journal} {\bibinfo  {journal} {Phys. Rev. B}\ }\textbf {\bibinfo {volume}
  {105}},\ \bibinfo {pages} {115305} (\bibinfo {year} {2022})}\BibitemShut
  {NoStop}%
\end{thebibliography}%

\end{document}